\documentclass[times, 10pt,twocolumn]{article}
\usepackage{PaperFormat_latex}
\usepackage{times}

\usepackage{graphicx}        
\usepackage{subfigure}
\usepackage{amsmath}

\usepackage{fancyhdr}

\begin{document}

\title{An Efficient Authorship Protection Scheme for Shared Multimedia Content}


\author{
Mohamed El-Hadedy, Georgios Pitsilis and Svein J. Knapskog\\
\\
The Norwegian Center of Excellence for Quantifiable Quality of Service in Communication Systems \\ (Q2S)\\
Norwegian University of Science and Technology (NTNU)\\
O.S.Bragstads plass 2E, N-7491 Trondheim, Norway
\email{hadedy@alumni.ntnu.no,georgios.pitsilis@gmail.com}
}

\maketitle
\thispagestyle{fancy}
\fancyhf{}
\renewcommand{\headrulewidth}{0pt}
\fancyfoot[L]{ICIG 2011, 6th International Conf. on Image and Graphics, Aug. 2011, ISBN: 978-0-7695-4541-7, IEEE Computer Society}

\begin{abstract}

Many  electronic content providers today like  Flickr and Google, offer space  to users to publish their electronic  media (e.g. photos and videos) in  their cloud infrastructures, so that  they can be publicly accessed.  Features like including other  information, such as keywords or owner  information into the digital  material is already offered by existing  providers. Despite the useful features made available to users by such  infrastructures, the authorship of  the published content is not  protected against various attacks such as  compression. In this paper we  propose a robust scheme that uses digital  invisible watermarking and  hashing to protect the authorship of the  digital content and provide  resistance against malicious manipulation of  multimedia content. The  scheme is enhanced by an algorithm called MMBEC, that is an extension of  an established scheme MBEC, towards higher resistance.

\end{abstract}

\section{Introduction}
Public  services,  like Flickr \cite{15}, Google \cite{16} and Instagram \cite{20}, which offer space to users to publish  their  electronic media and making them accessible to other people, have achieved high popularity today.

Information  such  as the owner of the photo as well as other information related  to  technical characteristics can already be captured and saved along  with  the media to help indexing and searching. Keywords that describe  the  content of photos submitted by users can yet be attached as  meta-data,  adding in this way semantic meaning to the content. In current  implementations of such services (e.g Flickr, Google) such  functionality  is already offered, but with all meta-data been stored  separately from the  content. Even though such solutions provide good  functionality, they do  not offer any added security, since no protection  of authorship is provided. This comes from the fact that, for public  content anyone is  given the reading permission and hence the  opportunity to copy,  republish and claim the authorship of it. In the  same way, owner  information or annotations that had been originally  associated to the  media by its owner also become vulnerable, as they can  be replaced either  mistakenly or maliciously.
Applying  ordinary digital watermarking onto the content would not provide  sufficient protection against the above threat. \emph{lossy compression},  \emph{geometric distortion} and \emph{addition of noise} or \emph{filtering} are ways to remove the watermarking and hence making unable to uniquely state  the authorship. On the other hand, \emph{visible watermarking}, while it is easy to apply, it requires significant intervention on the aesthetic view of the image.

We  propose  a new robust scheme which combines hash function and watermarking to  protect the authorship of the content against various  attacks. Compared  to existing robust solutions \cite{1} against  manipulation  of digital multimedia material, ours has the advantage of not requiring  public key encryption and therefore it is neither CPU intensive, nor  does it produce large volumes of data. It also uses blind  verification,  making the provision of the original object in the  verification  unnecessary, and therefore it is faster, easier and more secure. In  addition, in our proposed scheme, hashing is applied onto the whole  content thus being more reliable in the verification, as opposed to  hashing only part of the content.

The contribution of this paper is twofold. First, we introduce a novel  watermarking algorithm called MMBEC which is resistant to  various  attacks, and second, we propose a scheme which implements this  algorithm  and can be used for protecting the authorship of shared media content on the web. The rest of the paper is organized as follows: In section \ref{sec:RelWork} we state the problem and we present existing knowledge and related work in the field of watermarking. In section \ref{sec:PropScheme} we provide detailed description of the proposed scheme, while in section  \ref{sec:MMBEC} we present a short description of the proposed watermarking  algorithm. Experimental results  which demonstrate the ability of our  watermarking algorithm to resist compression attacks against other  alternative solutions are shown in section \ref{sec:Resul}. Finally our conclusions  and discussion follows in section \ref{sec:Concl}.

\section{Problem statement and Related Work}
\label{sec:RelWork}

The great ease in which digital images and data in general can be edited or duplicated has led to the need for effective tools for protecting their authorship. 
Digital Watermarking \cite{17} is known as one of the best solutions to prevent illegal duplication, redistribution and modification of digital multimedia. 
The process of watermarking regards the embedding of additional data along with the digital content prior to publishing the material to a Content Provider.

We consider \emph{Content Provider} as to be a service, mostly cloud-based, which provides public access to various types of material uploaded by the users. The aforementioned Flickr and Instagram, are two of the best known and most featured services of that kind.
The practice used today is that, the user who has performed the uploading of the digital material to the Content Provider is assumed to be the author of that material.
Nevertheless, for the reason that the content itself does not carry any evidence that could prove the identity of the real author, it makes easy to anyone to claim the authorship after the content has become publically available.
We challenge the idea of the authorship being embedded  into the material itself for the reason that in this way such evidence could be preserved over the lifetime of the object. That requires from the evidence to be embedded in a form that is resistant to manipulation.

Our idea for protecting the authorship of digital material is based on a simple principle:
Original multimedia content (e.g a digital photo) should contain watermarks only by the real author, which would be preserved over any subsequent attempts for re-watermarking.

The existence of an instance of the digital content carrying the identity of a single author would suffice to prove that this author is the real owner of the media. For the case that a user claims the authorship of the watermarked image, a protection scheme should provide a means of resolution carried out by a Trusted Third  Party (TTP), which can verify the existence of the identity of the real author in the media. In a real system such task can certainly be carried out by the Content Provider itself.

We should note that, it is most likely that the potential attackers would not be aware of the existence of watermarks in the media they attempt to attack. That is because, in the way watermarking is carried out, there is no way to confirm the existence of watermarks, if not knowing the secret keys used.

\begin{figure*}
\centering
\includegraphics[width=165mm]{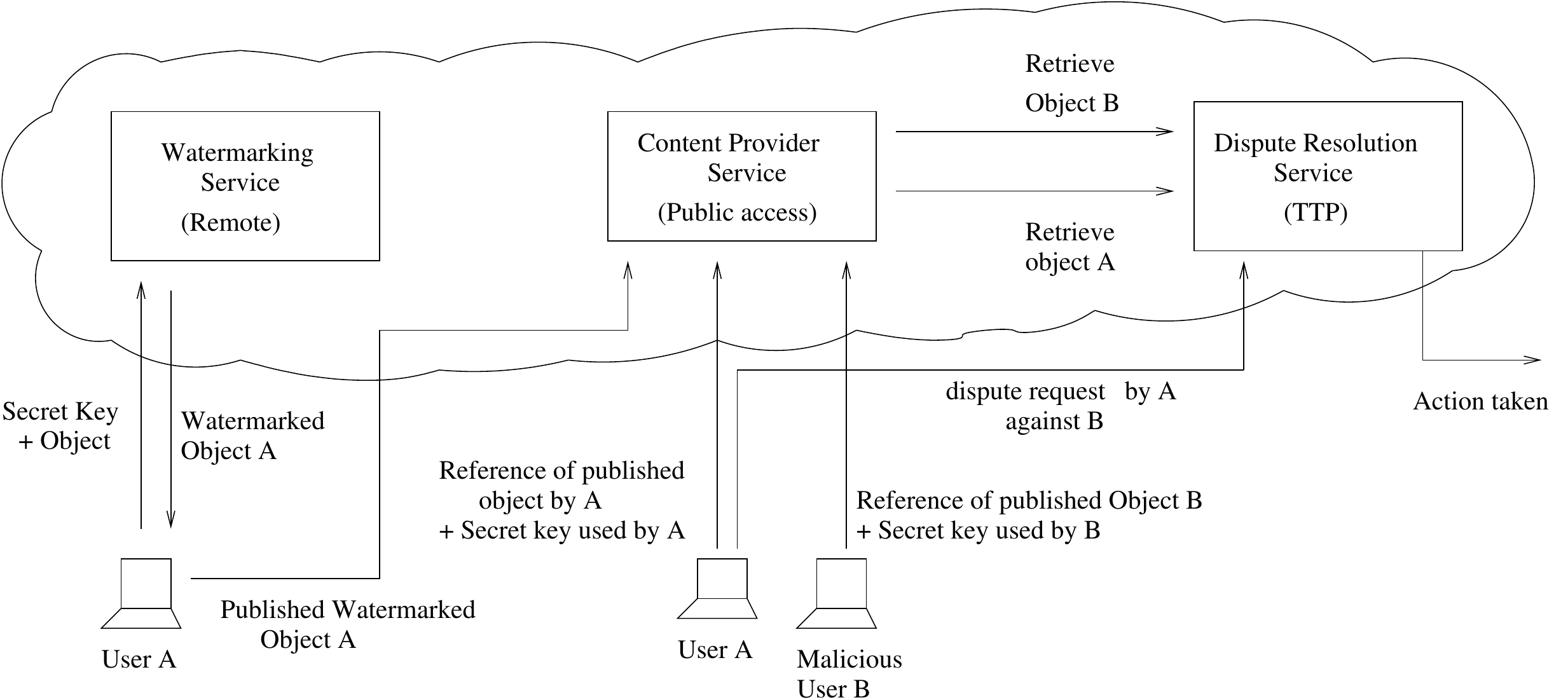}
\caption{System Architecture}
\label{fig:watermarking}
\end{figure*}

Next we describe two likely scenarios, depicted in fig.\ref{fig:watermarking}
of the authorship been attacked, and we show how the attempt should be overcome by the use of our proposed architecture.

\vspace{0.2cm}
\textbf{Scenario 1: \emph{The image is 'Re-published' by the attacker in its retrieved form}}

First, user \emph{A} wishes to publish a digital image to a public Content Provider service of his liking. The user, in order to  protect  his authorship, would watermark 
the  object prior to uploading it, using a secret key which has the form of text string. The watermarking process can be carried out either locally at the user's  end, or by using a trusted public watermarking service. For the latter case we assume the existence of secure communication channel between the service and the user. Once the content has been watermarked, user \emph{A} proceeds to publishing the watermarked image to the Content Provider service. 
Later, a malicious user \emph{B} retrieves from the Content Provider the published image \emph{A} onto which he attempts modifications. Finally he republishes the modified content at a Content Provider (either the same or a different from the one used by \emph{A}). From then on, any conflicting claims of authorship on this object should made possible to resolve via a verification process. Verification regards the checking of the existence of the embedded watermark into the published object. For example, if the material is found to be published elsewhere, it should be verifiable whether that was done without the permission of the real author.

\vspace{0.2cm}
\textbf{Scenario 2: \emph{The image is 'Re-watermarked' before it is 'Re-published' by the attacker}}

In  a more complex scenario, the attacker \emph{B} might also use the watermarking service to re-watermark the stolen image, for embedding his own secret text string in it, either knowing whether the original image is watermarked or not. In the case of dispute, either by the original user \emph{A}, or by the attacker \emph{B} claiming the authorship of the image, the verification service should be able to distinguish the real author by verifying and  confirming the existence of the hidden watermarks. In the above example it would not be possible for the attacker \emph{B} to prove that he is the author of the original image, since the existence of his watermark can not be confirmed in the watermarked copy published by the original author.  Meanwhile both watermarks (the real author's and the attacker's) would appear to have been embedded into the image published by the attacker. The existence of a copy of the image not containing the attacker's identity should be regarded as sufficient evidence to prove that he/she is not the real owner of the resource.

We summarize our assumptions into the following:

\begin{enumerate}
\vspace{-0.1cm}

\item A potential attacker is most likely not aware of the existence of watermarks in the digital media that he attempts to attack.

\item Re-watermarking on the same digital media can be attempted more than once. Each attempt does not remove the likely existing watermarks from the media.

\item Any attempt to attack an invisibly watermarked image by removing the hidden identity from it would severely degrade the underlying data, rendering the resulting image useless.

\item There exists a Trusted Third Party (TTP) acting as \emph{Dispute Resolution Service} which, with the provision of the appropriate evidence, can confirm the existence of a watermark on an image.
\vspace{-0.1cm}
\end {enumerate}

Using watermarks for protecting the copyright in various forms of digital media has been the subject of the research community for long time. We can distinguish two possible approaches in the use of watermarks, \emph{a)} the \emph{asymmetric watermarking schemes}, which require a pair of secret/public keys, and \emph{b)} the use of \emph{zero-knowledge proof protocols}.

The work by Herrigel et at. in \cite{18} is a characteristic example of watemarking technique for images, which belongs to the first category of approaches. Their technique provides robustness against various attacks with the advantage of not needing the original cover-image for the watermark detection. Contrary to our scheme, their method uses public-key cryptography and requires PKI infrastructure for supporting the distribution of public keys between the involved parties for mutual authentication. PKI infrastructure has also been found necessary to other solutions \cite{1}, used for video watermarking. Our approach has the advantage of neither needing the cover image to be provided in the verification, avoiding any exposure to other parties, nor does it need a PKI to be set up. 

It is known that watermarking-based approaches that make use of \emph{Zero knowledge protocols} must involve a TTP that take part in the embedding phase. Such solution has been reported to have potential drawbacks \cite{22}. In \cite{19}, Adelsbach et al. propose an approach based on the use of \emph{Zero Knowledge Proof}. Their protocol allows a prover to convince a verifier of the presence of a watermark without revealing any information that the verifier could use to remove the watermark, something known as \emph{blind watermark detection}. Nevertheless, their protocols still require the existence of another arbitrary Third party along with a Registration Center entity. On the contrary, our approach, having the advantages of \emph{blind detection}, it uses a much simpler but efficient protocol that employs one-way functions for achieving the same result. 

In this work we focus on the protocols used for embedding the watermarks into images and for verifying their existence.

\section{Proposed Scheme}
\label{sec:PropScheme}

Our proposed scheme uses hash functions and it comprises two phases: \emph{Embedding} and \emph{Verification}.  We  present a high level view which describes our previous scenario  of  uploading a multimedia object (e.g a photo) to a content provider  and  verifying its authorship. Furthermore, attacks based on modifying  already watermarked images (e.g jpeg compression) and re-watermarking  them can  be overcome by the use of a compression-resistant algorithm  such as the one we introduce in this paper.

\subsection{Embedding}

\begin{figure}[!t]
\centering
\includegraphics[scale=0.38]{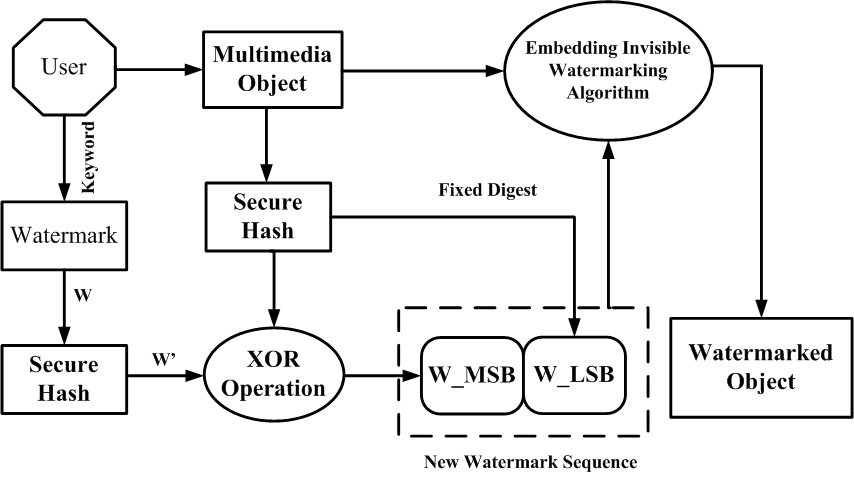}
\caption{Embedding Architecture scheme}
\label{fig_emb}
\end{figure}

The process of embedding is shown in Fig. \ref{fig_emb} and it consists of four main steps:

1.  The  user chooses its own secret keyword which is hashed. The reason for  hashing is to produce a digest which is a unique fixed length sequence  to be used as the  watermark. That gives the advantage of using secret  keywords of any length. There are several hash functions which can be  used. We refer to SHA-1  and SHA-2 \cite{3} as the simplest and best known.

2.  Along  with step 1, the multimedia object (e.g a JPEG,BMP,GIF image) is  also  hashed producing a fixed length digest of the object. The main reason  for doing this is to create a unique sequence for that object. Hash  functions  used in the previous step can also be used here.

3. The hashed watermark value is XOR-ed with the hashed object digest to produce the most significant part (\emph{W-MSB}) in the new watermark sequence as shown in Fig. \ref{fig_emb}. On the other hand, the hashed object digest make up the least significant part (\emph{W-LSB}) of the watermark sequence. That part is used as decryption key in the verification phase.

4.  The  new watermark sequence is then hidden inside the multimedia object  using a blind invisible watermarking technique such as MMBEC (Modified  Mid-band Exchange Coefficient) we are proposing. This algorithm is  described in the following section and it is shown that it has a high  level of robustness against normal attacks, such as compression. 

The plain object along with the secret keyword can be uploaded to a trusted  watermarking service provider for being watermarked.
Alternatively, the \emph{Embedding} can be carried out at the user's end before the uploading of the material, using trusted software provided by the Content provider.

\subsection{Verification}
\vspace{-0.2cm}

\begin{figure}[!t]
\centering
\includegraphics[scale=0.40]{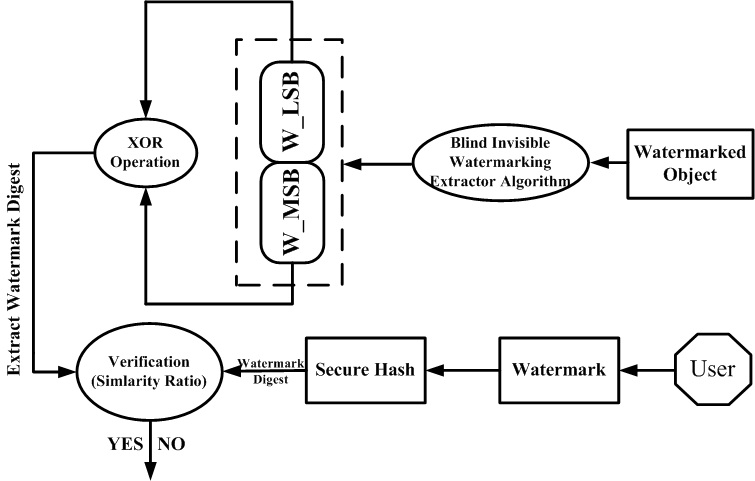}
\caption{Verification Architecture scheme}
\label{fig_ver}
\end{figure}

This phase is used for verifying the authorship of the content and it is shown in Fig. \ref{fig_ver}.  In  our design the verification can be carried out by the content provider  itself, by whom the identity of the real author can be distinguished  without knowing the original content.
The  verification requires as input the watermarked image along with the hidden keyword which is claimed having been embedded by its creator. As noted, we make the assumption that the hidden keyword is send through a secure communication channel to the content provider during the verification. Using a Secure Sockets Layer (SSL) protocol would suffice. The verification process is as follows:

1. The  watermarked secret keyword is hashed producing a hashed watermark  digest. Beside this, by applying blind invisible watermarking extraction  onto the watermarked object both the \emph{W-MSB} and \emph{W-LSB} digests can be retrieved. 

2. The \emph{Extracted Watermark Digest} of the object is obtained by XOR-ing the \emph{W-MSB} and \emph{W-LSB}.

3. Finally, the watermark digest is checked with respect to similarity to the extracted watermarked digest.

In  a real use case, a user who claims the authorship of an object which  has been modified and republished by an attacker (in some form we call a  \emph{Fake} object), can raise a dispute with the Content provider.
The Content provider in response, acting as TTP would verify both the claimant's and the attacker's hidden keywords against the \emph{Original} and \emph{Fake} objects. The \emph{Fake} object  is detected in the verification since the presence of both watermarks  (original users' and attacker's) may be confirmed in the attacker's  object, while the attacker's watermark is not confirmed in the original  user's object. A high similarity value between the extracted watermark  digest and the watermark digest of the secret keyword should be enough  to confirm the existence of a watermark into the verified object. It  should be noted that the original object should be kept secret by the  original author, who should publish to the Content provider the  watermarked version only. In this way the original image is not exposed  to the attackers.
The idea for the  Content provider to be acting as TTP and carry out the verification  process was made for the reason to protect the secrecy of the  watermarking keywords from the potential attackers.

\section{The MMBEC watermarking algorithm}
\label{sec:MMBEC}
Watermarking  algorithms  are classified in various ways. With regard to the  transformation they  do on the original image they are divided into \emph{Spatial Domain} and \emph{Frequency Domain} algorithms.  The former entail transformation of image pixels and is easy to  apply  to any image. The later perform transformation in the frequency  domain  and are more robust \cite{5}.
Whether the watermark information can be directly perceived by humans or not, watermarking is classified into: \emph{Visible} and \emph{Invisible}.
In the latter category the best known transformations used for hiding the watermark are \emph{Discrete Wavelet Transform} (DWT) and \emph{Discrete Cosine Transform} (DCT).

MBEC  is a known technique which utilizes the values of the mid-band  DCT  coefficients to encode a single watermark bit into a DCT $ 8 \times 8 $  block \cite{6}.
Our  proposed algorithm MMBEC, is an extension of the classical MBEC  algorithm. The purpose of creating a new algorithm is to achieve the  increased robustness required by the verification process. With  robustness we mean the sensitivity of the algorithm in identifying the  watermarks claimed to have been embedded into a digital object. 

\begin{figure}[!t]
\centering
\includegraphics[scale=0.45]{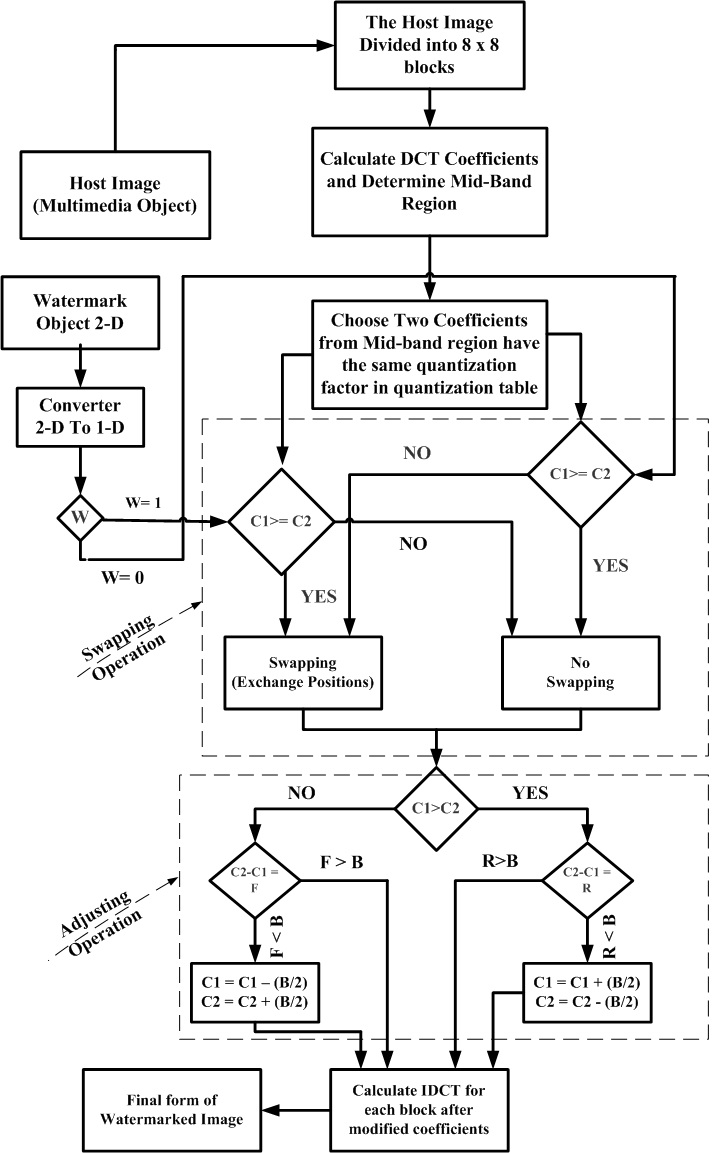}
\caption{Embedding operation in MMBEC}
\label{fig_n}
\end{figure}

MMBEC introduces two additional steps over the existing MBEC  as shown in Fig.\ref{fig_n} surrounded by dotted line, and which are described as follows:

\begin{enumerate}
\vspace{-0.1cm}
\item Choosing the exchange coefficients in Mid-band region according to the Quantization factors - shown in the top of Fig. \ref{fig_n}
\vspace{-0.2cm}
\item Increasing  the  difference between the selected coefficients to increase  the  robustness against several attacks by using strength factor B. 
\vspace{-0.1cm}
\end {enumerate}

\subsection{The embedding operation in MMBEC algorithm}

The watermark can be considered as a two-dimensional vector of binary digits. 
The aforementioned step 1 can further be divided into the following two operations:

\begin{enumerate}
\vspace{-0.1cm}
\item    The host image is divided into $ 8 \times 8 $ blocks, on which DCT transformation is applied afterwards. 
\vspace{-0.1cm}
\item    The watermark is converted from a two-dimensional vector into a one-dimensional sequence vector $W_i$, where $i=1,2,...,N \times M$. $N,M$ denote as the length and width of the watermark.
\vspace{-0.1cm}
\end {enumerate}

\begin{table}
\caption{Quantization values used in JPEG compression scheme}
\label{fig_sx}
\centering

\vspace{-0.3cm}
\begin{tabular}{|c|c|c|c|c|c|c|c|}

\hline
16 & 11 & 10 & 16 & 24 & 40 & 51 & 61 \\
 \hline
12 & 12 & 14 & 19 & 26 & 58 & 60 & 55 \\
\hline
14 & 13 & 16 & 24 & 40 & 57 & 69 & 56 \\
\hline
14 & 17 & 22 & 29 & 51 & 87 & 80 & 62 \\
\hline
18 & 22 & 37 & 56 & 68 & 109 & 103 & 77 \\
\hline
24 & 35 & 55 & 64 & 81 & 104 & 113 & 92 \\
\hline
49 & 64 & 78 & 87 & 103 & 121 & 120 & 101 \\
\hline
72 & 92 & 95 & 98 & 112 & 100 & 103 & 99 \\
\hline
\end{tabular}
\vspace{-0.2cm}

\end{table}

Two coefficients $C1_b(x, y)$, $C2_b(x, y)$ in each $8 \times 8$ block from the middle-band are chosen from the quantization $8 \times 8$ table for $jpeg$ in Table \ref{fig_sx}. The use of quantization tables is introduced in \cite{11}.
The  watermark will be divided into smaller parts in the compression  operation for being inserted into the mid-band region. The perfect  locations for insertion are those which the quantization factor in the  table has the values as follows:

\begin{itemize}
\vspace{-0.2cm}
\item  The coefficients $C1_b(x, y)$ and $C2_b(x, y)$ are located in positions (4, 1) and (2, 3) and quantized by a factor equal to 14.
\vspace{-0.2cm}
\item  The coefficients $C1_b(x, y)$ and $C2_b(x, y)$ are located in positions (3, 3) and (1, 4) and quantized by a factor equal to 16.
\vspace{-0.2cm}
\item  The coefficients $C1_b(x, y)$ and $C2_b(x, y)$ are located in positions (5, 2) and (4, 3) and quantized by a factor equal to 22.
\vspace{-0.2cm}
\item  The coefficients $C1_b(x, y)$ and $C2_b(x, y)$ are located in positions (3, 4) and (1, 5) and quantized by a factor equal to 24.
\vspace{-0.2cm}
\item  The coefficients $C1_b(x, y)$ and $C2_b(x, y)$ are located in positions (3, 5) and (1, 6) and quantized by a factor equal to 40.
\vspace{-0.2cm}
\end{itemize}

Next, step 2 of the MMBEC algorithm includes \emph{Swapping} and \emph{Adjusting} which are described as follows:

\emph{Swapping} is done similarly as in the original MBEC with the purpose of hiding the watermark.
It requires the selection of two suitable coefficients $C1_b$ and $C2_b$ to be swapped when either of the following two
 conditions is met:

\begin{itemize}
\vspace{-0.2cm}
\item $(W_i = 1)$ and $(C1_b (x, y) < C2_b (x, y))$. 
\vspace{-0.2cm}
\item $(W_i = 0)$ and $(C1_b (x, y) < C2_b (x, y))$. 
\vspace{-0.2cm}
\end{itemize}
$W_i$ denotes one bit of the binary sequence.


\begin{figure}
\centering
\includegraphics[scale=0.40]{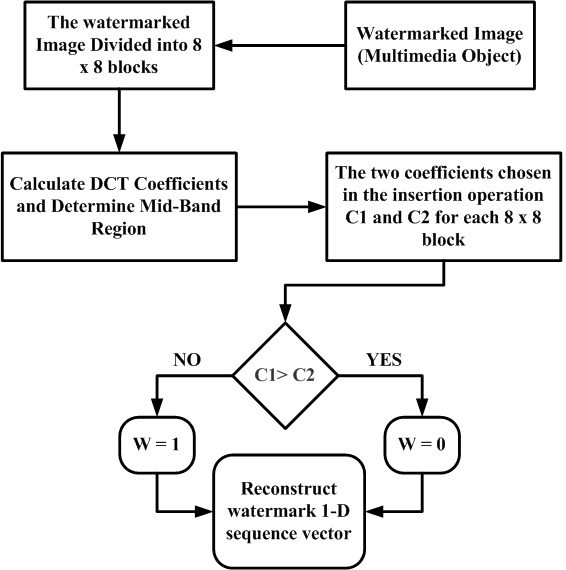}
\caption{Extraction operation in MMBEC}
\label{fig_ex}
\end{figure}


\emph{Adjusting} regards the process of increasing the robustness of the watermark and it is done by adding a strength factor $B$. Increasing $B$ reduces the chance of the watermark being extracted wrongly at the expense of additional image degradation.
The method of \emph{Adjusting} is shown below:
\begin{itemize}
\item  If ($C1_b(x, y) > C2_b(x, y)$)  and  ($C1_b(x, y)- C2_b(x, y) < B$)
\\ Then 
$
\begin{cases}
C1_b(x, y) \leftarrow C1_b(x, y) + \frac{B}{2} \\
C2_b(x, y) \leftarrow C2_b(x, y) - \frac{B}{2} \\
\end{cases}
$

\item  If ($C1_b(x, y) < C2_b(x, y)$)  and  ($C2_b(x, y)-C1_b(x, y) < B$)
\\ Then            
$
\begin{cases}
C2_b(x, y) \leftarrow C2_b(x, y) + \frac{B}{2} \\
C1_b(x, y) \leftarrow C1_b(x, y) - \frac{B}{2} \\
\end{cases}
$

\end{itemize}

Finally, the watermarked image is reconstructed by applying the Inverse Discrete Cosine Transform (IDCT)\cite{10} to the output of \emph{Adjusting}.

\subsection{The watermark extraction in MMBEC algorithm}

For the watermark extraction the original image is not needed as shown in Fig.\ref{fig_ex}. Similarly as in embedding, for each block the values of $C1_b(x, y)$ and $C2_b(x,y)$ are compared with each other giving back the binary sequence of the watermark as follows:
\vspace{-0.1cm}
\begin{itemize}
\item     If    $C1_b(x, y) \geq C2_b(x, y)$ Then $W_i \leftarrow 0$ Else $W_i \leftarrow 1$
\end{itemize}

\section{Experimental Results}
\label{sec:Resul}

We present the results of performance tests and comparison of strength of our algorithm against compression attacks.

\begin{figure}
 \centering
 \subfigure[LENA]{
 \includegraphics[scale=0.25]{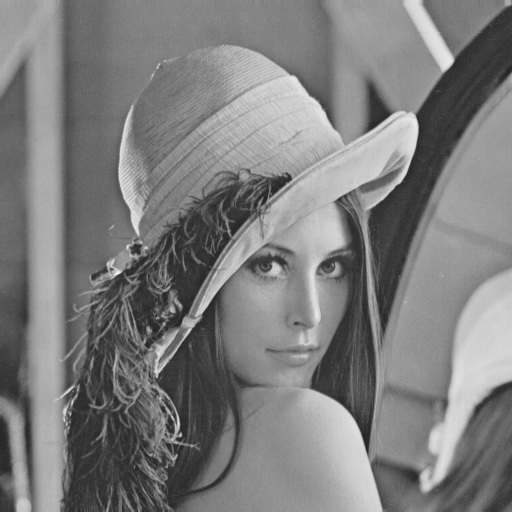}
 \label{fig:subfig1}
 }
 \subfigure[The Watermark]{
 \includegraphics[scale=0.4]{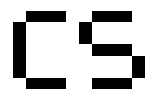}
 \label{fig:subfig2}
 }
 
 \label{fig:subfigureExample}
 \caption[The original Image(Host Image)]{The test images}
 \end{figure}

For the evaluation we used a gray-level image of $ 256 \times 256 $ resolution shown in Fig. \ref{fig:subfig1} in which we tried to embed the watermark object shown in Fig. \ref{fig:subfig2}.

The level of \emph{JPEG} compression applied to the image is expressed by the \emph{Quality Factor} (QF).  The  optimal value is dependent on both the level of detail contained in the  image and the already applied compression ratio. In our evaluation, the  QF ranged from 5 to 100.
In  practice, values of QF between 0 and 24 correspond to the extreme case  of applying  high compression to the image. Since  the details of  the  picture as well as the hidden watermark are severely distorted when  applying such levels of compression, we consider such cases to be  unrealistic.
We used a metric we called \emph{Similarity Ratio} (SR) to demonstrate the effect of compression to the retrieval of the watermark during the verification.  
This  relates to the similarity between the elements of the binary sequence  vector of the Extracted Watermark Digest and the digest of the secret  word provided as input to the verification.
SR is given in formula \ref{SimilarityRatio}, where S denote as the number of matching pixel values between the pictures of the two digests mentioned above,  
and D the number of different pixels they have.

\vspace{-0.1cm}
\begin{equation}
SR = S \cdot (S+D)^{-1}
\label{SimilarityRatio}
\end{equation}

\begin{figure*}
\centering
\includegraphics[width=1.0\textwidth, height = 2.45 in]{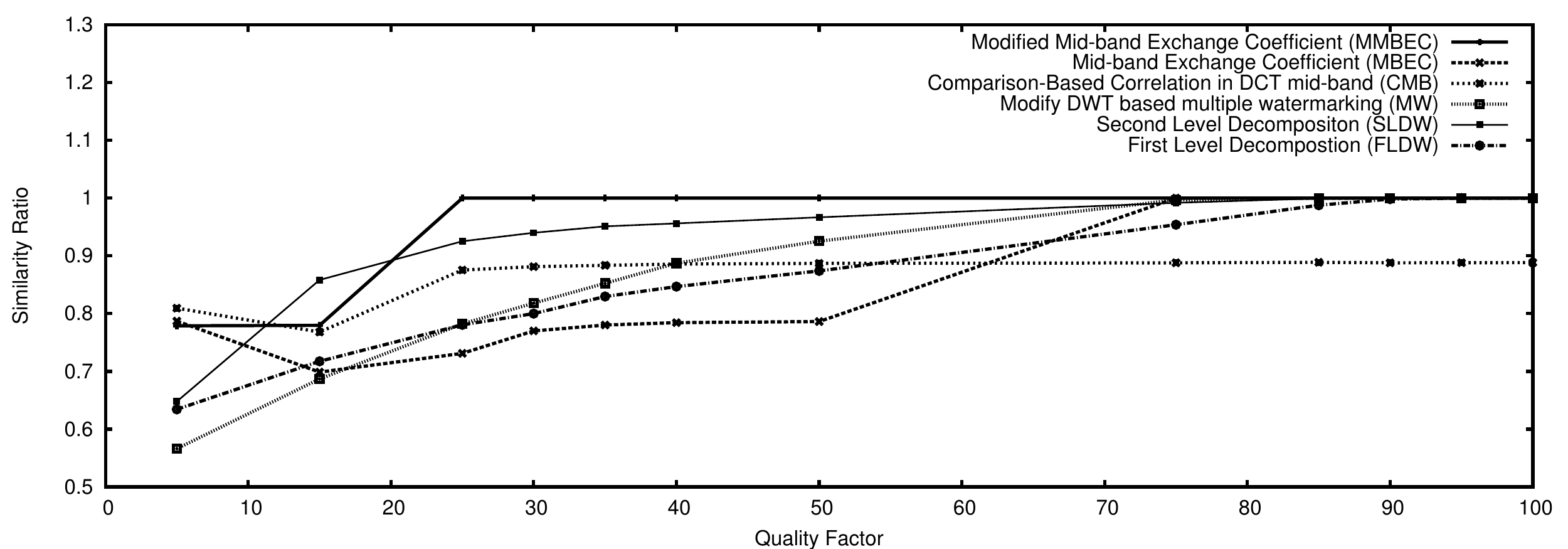}
\caption{The relationship between SR and QF}
\label{fig_vp}
\end{figure*}

In Fig.\ref{fig_vp} is  shown  the comparison diagram between the MMBEC algorithm and five other algorithms used for embedding and extracting the watermark from  the same original test image. In table \ref{fig_num} we present the numerical values of this comparison. 
The symbols used for each technique are:
 SR1:(MMBEC) for our proposed technique, SR4:Modified DWT based  multiple  watermarking (MW)\cite{7}, SR5:Second Level Decomposition Wavelet (SLDW)\cite{8}, SR6:First Level Decomposition Wavelet (FLDW)\cite{8}, SR3:Comparison-based correlation in DCT mid-band (CMB)\cite{9} and SR2:Mid-band Exchange Coefficient (MBEC)\cite{6}. The diagram demonstrates how the effect of compression develops over various values of QF.
Very  interestingly, as can be seen in the figure, both MMBEC and CMB  algorithms behave very steadily when QF remains  within the range of 25  and 100, with MMBEC achieving higher Similarity Ratios. Meanwhile for  MBEC, MW, FLDW and SLDW there  is an increasing trend for the same range  of QF. In addition, with MMBEC the highest possible resistance against  compression ($SR = 1$) can be achieved for images not compressed very much ($QF>25$).  Moreover,  no other algorithm could reach such good performance for such  a wide  range of compression (QF) in the way MMBEC does.

We can attribute the reason for the high performance achieved by MMBEC to the way that the two coefficients $C1_b(x, y)$, $C2_b(x, y)$ are chosen from the quantization table for hiding the image in the mid-band region.

\begin{table*}
\centering
\caption{The relationship between SR (Similarity Ratio) and QF (Quality Factor) (Numerical values)}
\label{fig_num}
\centering

\begin{tabular}{|c||c|c|c|c|c|c|}
\hline
 QF   & SR1 	& SR2	& SR3	& SR4	& SR5	& SR6 \\
\hline
\hline
5    & 0.7790  & 0.7871     & 0.8094	& 0.5662 &	0.6479 &	0.6343 \\
 \hline
15   & 0.7797	& 0.6987    & 0.7681    & 0.6868 &	0.8583 &	0.7175 \\
\hline
25   & 1.0000	& 0.7313	& 0.8754	& 0.7823 & 	0.9252 & 	0.7807 \\
\hline
30   & 1.0000	& 0.7700	& 0.8814	& 0.8181 & 	0.9399 & 	0.8001 \\
\hline
35   & 1.0000	& 0.7802	& 0.8835 	& 0.8524 & 	0.9511 & 	0.8297 \\
\hline
40   & 1.0000	& 0.7844	& 0.8859	& 0.8873 & 	0.9562 & 	0.8468 \\
\hline
50   & 1.0000	& 0.7863	& 0.8870	& 0.9255 & 	0.9668 & 	0.8740 \\
\hline
75   & 1.0000	& 1.0000	& 0.8879	& 0.9968 & 	0.9919 & 	0.9539 \\
\hline
85   & 1.0000	& 1.0000	& 0.8889	& 1.0000 & 	0.9995 & 	0.9877 \\
\hline
90   & 1.0000	& 1.0000	& 0.8879	& 1.0000 & 	1.0000 & 	0.9982 \\
\hline 
95   & 1.0000	& 1.0000	& 0.8882	& 1.0000 & 	1.0000 & 	1.0000 \\
\hline
100  &  1.0000	& 1.0000	& 0.8883	& 1.0000 & 	1.0000 & 	1.0000 \\
\hline
\end{tabular}

\end{table*}

\section{Conclusion}
\label{sec:Concl}
Authorship protection of multimedia material has always been a concern of the  scientific community as well as of the simple users. We believe a proper  solution to this  problem would encourage more the online distribution of multimedia  material within user communities.

In  this  paper we proposed a scheme that can be applied on existing and future cloud infrastructures for protecting the authorship of digital  multimedia  material. The employment of invisible watermarking  algorithms for hiding  the user's identity in the watermarked image is  one of the strong points of our scheme. A Trusted Third Party is  involved in the task of resolving the identity of the real author of a  multimedia object, a role that the Content provider can certainly serve.
A  simple  to deploy and thus suitable for our use-case  attack-resistant  watermarking algorithm called MMBEC has also been  introduced in this  paper. The requirements we have originally set for  such an algorithm of being resistant against malicious modifications  performed on images, such as compression, are actually fulfilled, as  shown in our evaluation results.  Further testing its resistance against other  types of attacks is left for future  work.


\begin{thebibliography}{1}

\bibitem{1} Q.~Sun and D.~He and Q.~Tian, \textit{"A Secure and Robust Authentication Scheme for Video Transcoding"},IEEE transactions on circuits and systems for video technology, vol. 16, no. 10, October 2006.

\bibitem{3} W. E. Burr, \textit{"Cryptographic Hash Standards: Where Do We Go from Here?"}, IEEE Security and Privacy, Vol.4/2, pp.88-91, Mar./Apr. 2006.


\bibitem{5} B. Richart, R. Mart, J. Delgado, A. H. Sadka and P. Sweeney, \textit{"Definition of protocols for secure anonymous purchase"}, In proceeding of International Conference on Networking and Services (ICNS 2005), Tahiti, French ,October 2005.
\bibitem{10} Wen-Hsiung, C., C. Smith, et al. (1977). \textit{"A Fast Computational Algorithm for the Discrete Cosine Transform."} Communications, IEEE Transactions on 25(9): 1004-1009.
\bibitem{6} E. Koch and J. Zhao, \textit{"Towards robust and hidden image copyright  labeling"},In proceedings of IEEE Workshop on Nonlinear Signal and Image Processing, pp. 452—455, Neos  Marmaras, Greece, June 1995. 
\bibitem{7} R. Mehul S, R. Priti P, \textit{“Discrete wavelet transform based multiple watermarking scheme”}, IEEE TENCON 2003, Bangalore, India, Oct.2003.
\bibitem{8} D. Kundur, \textit{“Multiresolution digital watermarking: algorithms and implications for multimedia signals”}, doctoral thesis, Univ. of Toronto, 1999.
\bibitem{9} G.C. Langelaar, I. Setyawan, R.L. Lagendijk, \textit{"Watermarking Digital Image and Video Data: A State-of-the-Art Overview"}, IEEE Signal Processing Magazine, vol. 17, no. 5, pp. 20-46, September 2000.
\bibitem{11} J. D. Kornblum, \textit{"Using JPEG quantization tables to identify imagery processed by software"}, In proceedings of the Eighth Annual DFRWS Conference, vol. 5, Supplement 1, issn = 1742-2876, pp. S21 - S25, 2008.
\bibitem{15} M.J. Huiskes, M.S. Lew, \textit{"The MIR Flickr Retrieval Evaluation"}, ACM International Conference on Multimedia Information Retrieval (MIR'08), Vancouver, Canada, 39-43, 2008.
\bibitem{16} Vic, Gundotra, \textit{"Google+: Communities and photos"}, Google Blog. Retrieved 6 December 2012.

\bibitem{17} V. Singh, \textit{"Digital Watermarking: A Tutorial"}, Multidisciplinary Journals in Science and Technology, Journal of Selected Areas in Telecommunications (JSAT),January edition 2011.

\bibitem{18} A. Herrigel, J. Ruanaidh, H. Petersen, S. Pereira, and T. Pun. \textit{"Secure copyright protection techniques for digital images"}, In Information Hiding, LNCS 1525, pp. 169,190. 1998.

\bibitem{19} Andre Adelsbach and Ahmad-reza Sadeghi. \textit{"Zero-Knowledge Watermark Detection and Proof of Ownership"}, 4th Int. Workshop on Info. Hiding, LNCS 2137:273–288, Springer Verlag, 2001. 

\bibitem{20} http://instagram.com, \textit{"Instagram online photo-sharing, video-sharing and social networking service"}

\bibitem{22} Raphael C.-W Phan and Huo-Chong Ling. \textit{"Flaws in generic watermarking protocols based on zero-knowledge proofs"}, Third international conference on Digital Watermarking (IWDW'04), Springer-Verlag, Berlin, Heidelberg, 184-191, 2004.

%




\end{thebibliography}
\end{document}